\documentclass[twocolumn,runningheads]{svjour2}
\journalname{Astrophysics and Space Science}
\smartqed  
\usepackage{graphicx}
\newcommand{\crat}{s$^{-1}$}
\begin{document}

\title{The complex X-ray spectrum of the isolated neutron star
  RBS1223\thanks{This work was supported in part by 
  the German DLR under grant 50 OR 0404. Based on observations obtained with
  XMM-Newton, an ESA   science mission, with instruments and contributions
  directly funded by ESA   Member States and NASA} 
}
\author{Axel D. Schwope \and
        Valeri Hambaryan \and 
        Frank Haberl \and
        Christian Motch}

\institute{A.~Schwope \& V.~Hambaryan \at
              Astrophysikalisches Institut Potsdam, An der Sternwarte 16, 14482
              Potsdam, Germany \\
              Tel.: +49-331-7499232\\
              Fax: +49-331-7499429\\
              \email{ASchwope at aip.de}           
           \and
           F.~Haberl \&  V.~Hambaryan \at
           Max-Plank Institute f\"ur Extraterrestrische Physik,
           Giessenbachstr., 85748 Garching, Germany
           \and
           Christian Motch \at
           Observatoire Astronomique, CNRS UMR 7550, 11 rue de l'Universit\'e, 67000
              Strasbourg, France
}

\date{Received: date / Accepted: date}

\maketitle

\begin{abstract}
We present a first analysis of a deep X-ray spectrum of the
    isolated neutron star RBS1223 obtained with XMM-Newton. Spectral data
    from four new monitoring observations in 2005/2006 were combined with
    archival observations obtained in 2003 and 2004 to form a spin-phase
    averaged spectrum containing 290000 EPIC-pn photons.
    This spectrum shows higher complexity than its predecessors,
    and can be parameterised with two Gaussian absorption lines superimposed on a
    blackbody. The line centers, $E_2 \simeq 2E_1$, could be
    regarded as supporting the cyclotron interpretation of the absorption
    features in a field $B \sim 4 \times 10^{13}$\,G. 
    The flux ratio of those lines does not support this interpretation. Hence,
    either feature might be of truly atomic origin. 
\end{abstract}

\section{Introduction}
\label{intro}
RBS1223 is one out of three XDINs (X-ray dim isolated neutron stars)  
found in the ROSAT Bright Survey (RBS),
an optical identification program of the more than 2000 X-ray sources
detected in the RASS with a count rate $>0.2$\,\crat\ at high galactic
latitude \cite{r1}. The initial discovery \cite{r2} was based on the soft
X-ray spectrum and the steep SED, $f_{\rm X}/f_{\rm opt} > 
10^4$. Follow-up Chandra observations revealed a periodically modulated signal
\cite{r3}, and HST-observations uncovered a candidate optical counterpart at
$m_{\rm 50CCD} = 28.56 \pm0.13$\,mag \cite{r4}. Initial observations with
XMM-Newton showed deviations from a Planckian energy distribution at energies
below 500 eV which could be described with a Gaussian absorption line and
interpreted as a cyclotron absorption 
line in a field of $(2-6) \times 10^{13}$\,G \cite{fwh03}. The large number of
photons collected with XMM-Newton also uncovered the true spin period of
10.31\,s. In \cite{asea05} the light curves in different energy bands were
modeled in terms of a photospheric cap model. The temperature structure 
was found to be roughly compatible with the crustal field models of
\cite{urm04}. The implied field structure deviates from a simple centered
dipole model. Spin-period 
changes appeared likely at that time thus stimulating  monitoring observations
with Chandra and XMM-Newton. The spin history could be unequivocally fixed
by a series of Chandra pointings \cite{kap05} and revealed a spin-down of the
star at a rate $\dot{P} = 1.120 \times 10^{-13}$\,s\,\crat. Under the
assumption that this is due to magnetic dipole torques, a characteristic age
of 1.5 Myr and a magnetic field strength of $3.4 \times 10^{13}$\,G were
inferred. Here we describe initial results of recent monitoring observations
with XMM-Newton which, due to the large number of accumulated photons,
besides accurate timing also allows spectroscopy with unprecedented accuracy.

\begin{table}[t]
\caption{XMM-Newton observations of RBS1223 analysed in this contribution}
\centering
\label{t:obs}
\begin{tabular}{rclrcc}
\hline\noalign{\smallskip}
Rev. & AO & Day & Exp. & Mode & Countrate \\
&     &     & [ks]     & &[\crat]\\
\tableheadseprule\noalign{\smallskip}
377 & 1& 2001 Dec 31 & 18.6 & SW & 2.49$^{\rm a}$\\
561 & 2& 2003 Jan 1 & 27.0 & FF& 2.53 \\
743 & 3& 2003 Dec 30 & 30.2 & FF& 2.57 \\
1015& 4& 2005 Jun 25 & 14.9 & FF& 2.56 \\
1016& 4& 2005 Jun 26 & 12.9 & FF& 2.57 \\
1025& 4& 2005 Jul 15 & 12.9 & FF& 2.56 \\
1115& 4& 2006 Jan 10 & 14.9 & FF& 2.56 \\
\noalign{\smallskip}\hline
\end{tabular}\\
\begin{flushleft}
-- Exposure time is given for EPIC-pn. \newline
-- Mean countrates are given for the whole XMM-Newton energy range (0.15 --
10\,keV)\newline
-- $^{\rm a}$ countrate from \cite{fwh03}
\end{flushleft}
\end{table}

\section{XMM-Newton observations of RBS1223}
\label{sec:1}
XMM-Newton observed RBS1223 several times between 2001 and 2006. The relevant
pointings for this contribution are listed in Tab.~\ref{t:obs}. All 
but the first, which was put in small window mode (SW),  were obtained in full
frame mode through the thin filter.  All observations were reduced 
in a homogeneous manner with a recent version of the SAS (V6.5). The
monitoring observations set constraints on the evolution of different
parameters of the star. 
The overall brightness did not change between 2001 and 2006 and 
at all occasions the star displayed its double-humped
light curves discussed in detail by \cite{asea05}. Also the amplitude of the
light curve remained unchanged. Hence, 
the only property which was shown to evolve so
far is the spin period \cite{kap05}. 

\begin{figure*}
\centering
\resizebox{0.32\hsize}{!}{\includegraphics[clip=]{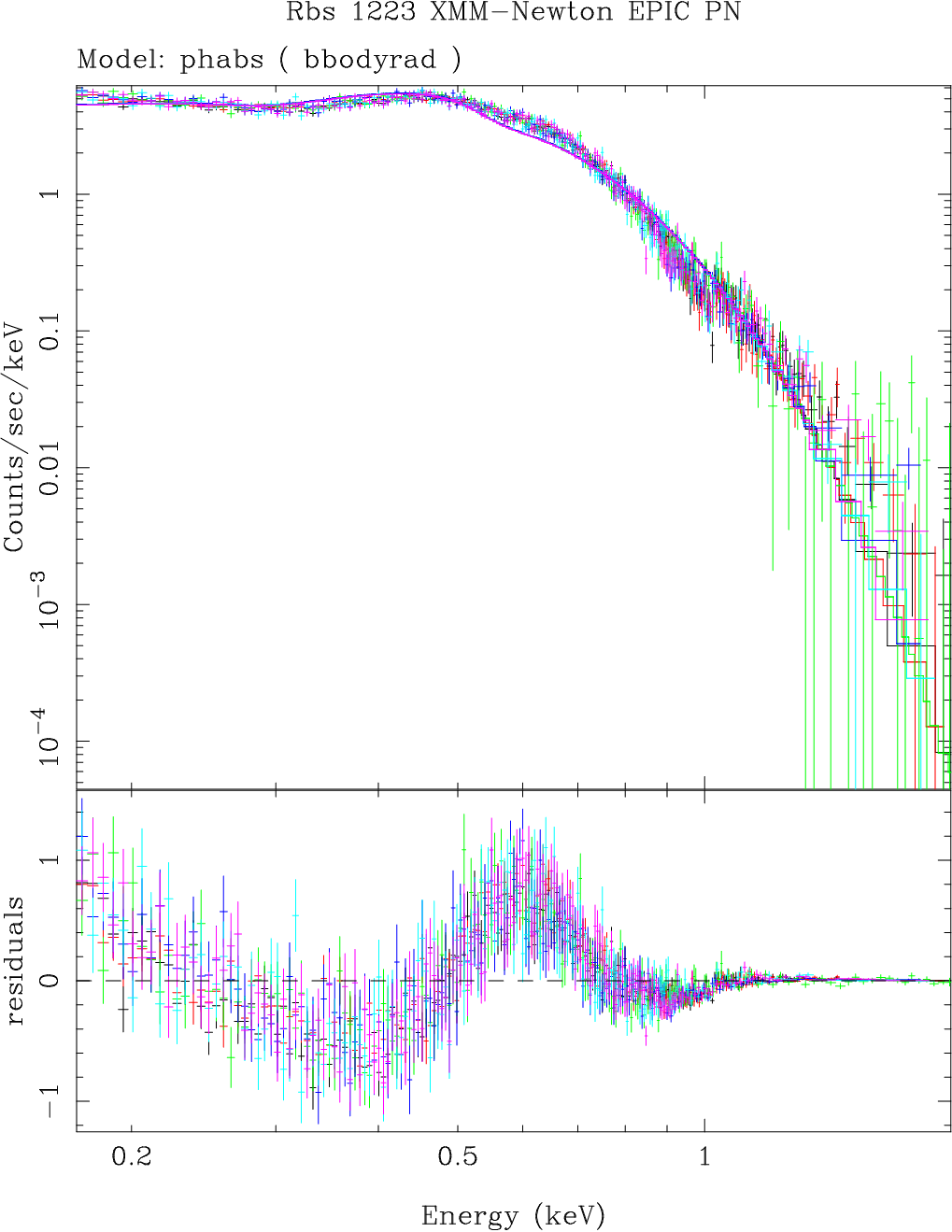}}
\hfill
\resizebox{0.32\hsize}{!}{\includegraphics[clip=]{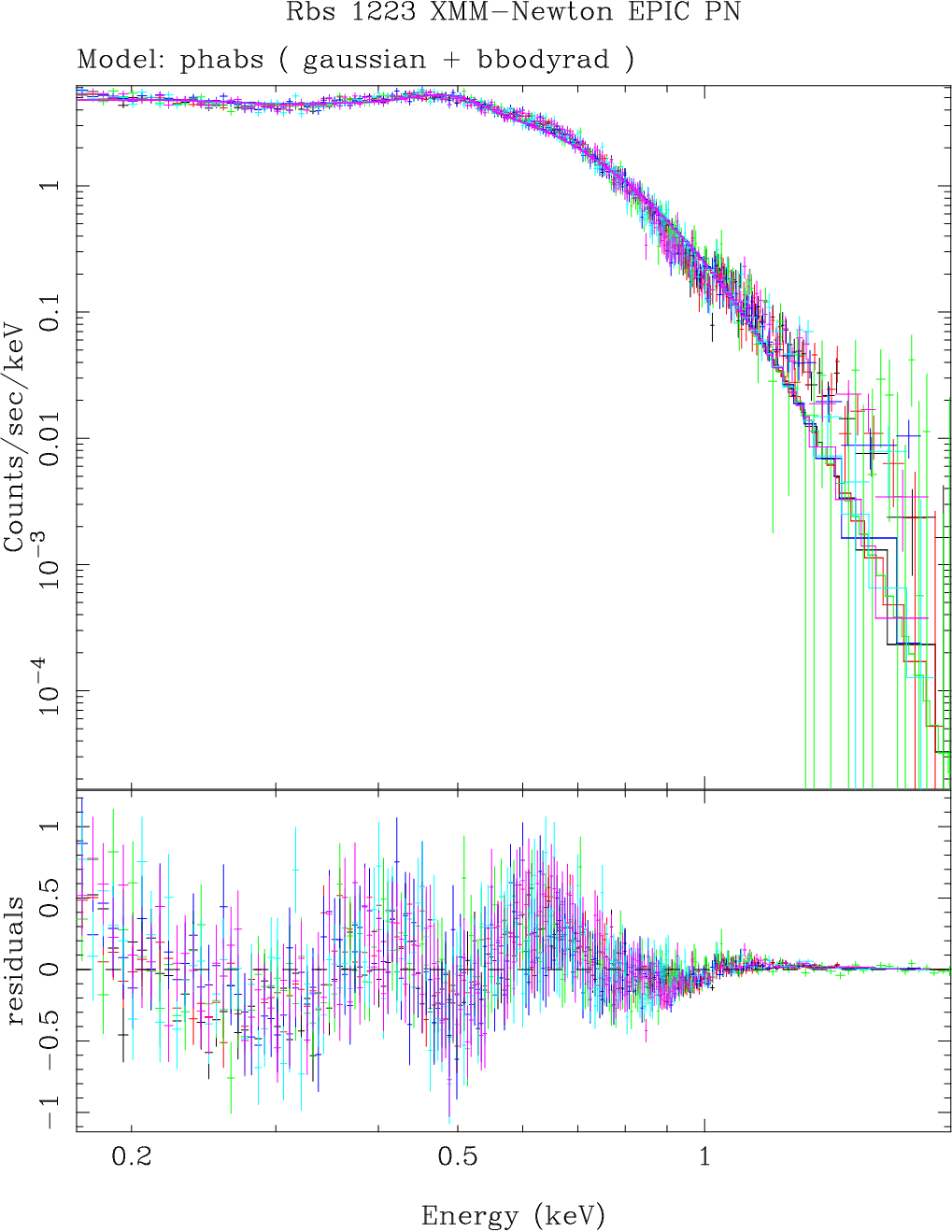}}
\hfill
\resizebox{0.32\hsize}{!}{\includegraphics[clip=]{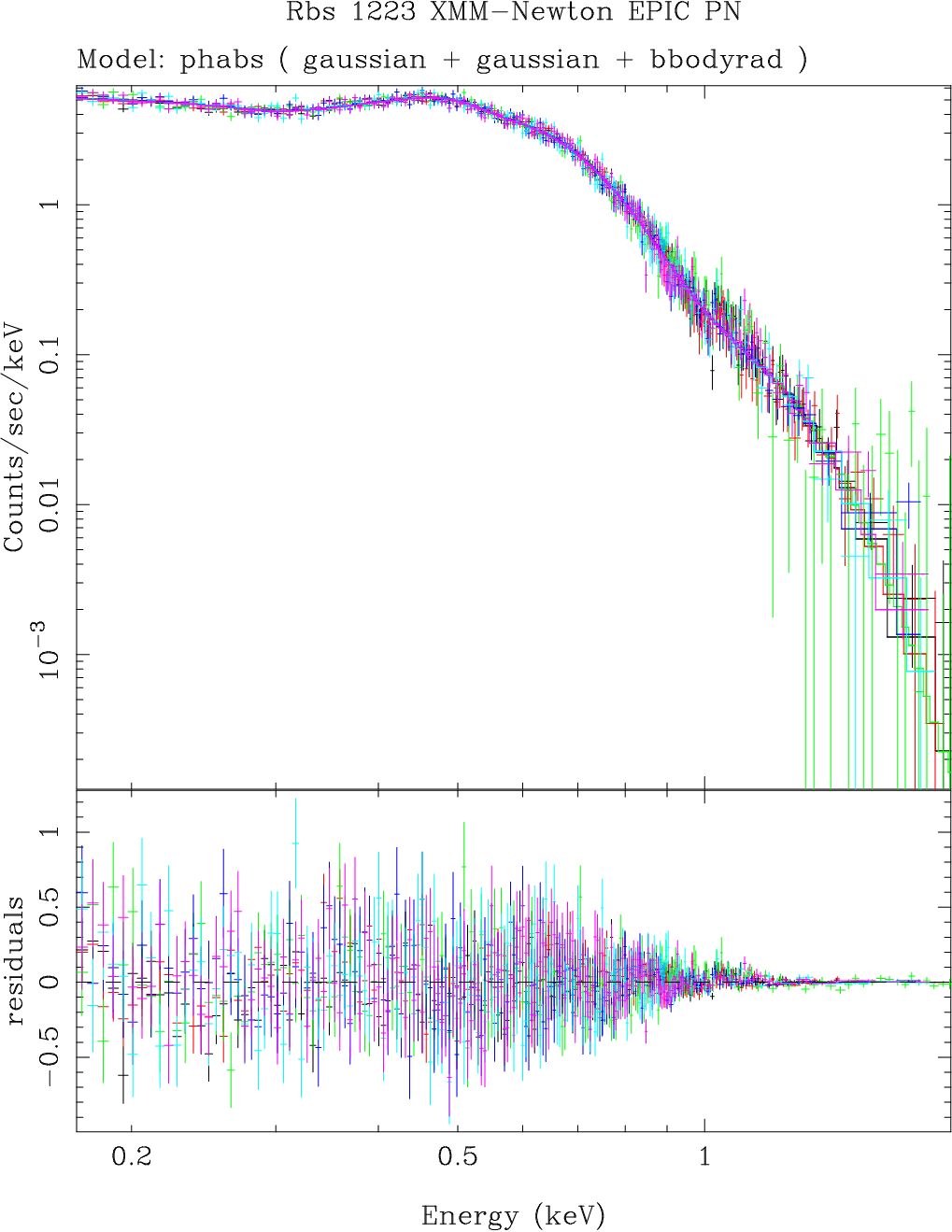}}
\caption{Results of spectral fitting ordered according to increasing
  complexity of the model. From
  left to right the models shown are an absorbed blackbody, a blackbody
  plus Gaussian absorption line, and a blackbody with two Gaussian absorption
  lines. The spectral parameters are listed in Tab.~\ref{t:spec}, models 1 -- 3}
\label{f:spec}
\end{figure*}
%
\begin{table*}[t]
\caption{Spectral modeling of RBS1223. $E$ and $\sigma$ denote the central
  energy and width of the two Gaussians (subscript $1$ and $2$), while $F$
  denotes the line flux}
\centering
\label{t:spec}       
\begin{tabular}{rccccccccrr}
\hline\noalign{\smallskip}
\# & $N_{\rm H}$ & $kT_{\rm bb}$ & $E_1$ & $\sigma_1$ & $F_1$ & $E_2$ &
$\sigma_2$ & $F_2$ & red.~$\chi^2$ & \#d.o.f\\
 & $10^{20}$ cm\,$^{-2}$ & eV & keV & keV & keV cm$^{-2}$ \crat & keV
 & keV & keV cm$^{-2}$ \crat & \\
\tableheadseprule\noalign{\smallskip}
1 & 3.3 & 100 &   --- & --- & --- & --- & --- & --- & 3.96 & 1250 \\
2 & 3.7 &  93 & 0.39 & 0.06 & $ -6 \times 10^{-4}$ & --- & --- & --- & 2.05 & 1247 \\
3 & 1.8 & 102 & 0.23 & 0.15 & $-35 \times 10^{-4}$ & 0.46 & 0.26 & $-7 \times 10^{-4}$ & 1.11 & 1244 \\
4 & 2.0 & 104 & 0.28 & 0.13 & $-21 \times 10^{-4}$ & $ = 2\times E_1$ & 0.24 & $-7\times 10^{-4}$ & 1.12 & 1245 \\
5 & 1.2 & 100 & 0.20 & 0.17 & $-27 \times 10^{-4}$ & 0.73 & $= \sigma_1$ &$-3\times 10^{-4}$ & 1.11 & 1245 \\
6 & 1.8 & 102 & 0.28 & 0.15 & $-35 \times 10^{-4}$ & $ = 1.5\times E_1$ & 0.26& $-7\times 10^{-4}$& 1.11 & 1245 \\
\noalign{\smallskip}\hline
\end{tabular}
\end{table*}

\section{Spectral analysis of RBS1223}
The constancy of the star allows a joint 
spectral analysis of all available observations. In order
to minimise possible remaining calibration uncertainties for different camera
modes we used only those observations with identical 
camera settings (six observations between revolutions 561 -- 1115).  This
selection left us with an effective exposure time of 113\,ks with
almost 290000 photons for spectral analysis. Here we concentrate on the mean
spectrum of all observations. Phase-resolved spectroscopy is underway and will
be presented elsewhere.  

An initial attempt to analyse the spectrum was made with a spectrum extracted
from a merged photon event table of the six observations and with common
response and effective area files. Such an approach does not take into
account possible time-dependencies of the detector files. We therefore
extracted a spectrum and computed the response matrix and effective area file
for each observation separately. Spectral modeling was performed with XSPEC
V12 following two different approaches. Firstly, the spectral
parameters were determined for each spectrum individually but forcing
the column density of cold interstellar absorbing matter,
$N_{\rm H}$, to be the same for all the observations (six data groups).  
This experiment revealed the spectral parameters of all
observations to be the same within the errors. We then followed the second
approach fitting all six spectra jointly with one
common model which reveals one set of spectral parameters instead of six. 
The results of those latter experiments are listed in Tab.~\ref{t:spec}. 

Spectral models were chosen with increasing complexity until a satisfactory
fit was achieved. The fit assuming only a blackbody emission spectrum
absorbed by neutral interstellar matter which gives $\chi^2_\nu \simeq 4$ (for
1247 d.o.f.) can be ruled 
out completely (model 1 in Tab.~\ref{t:spec}, Fig.~\ref{f:spec}a). This
mismatch was found  
already by \cite{fwh03} when fitting data from revolutions 377 \& 561. 
They successfully applied a model with a Gaussian absorption line superposed
on the absorbed blackbody. With such a model the current fit is improved
but it does no longer adequately represent the shape
of the combined spectrum (model 2 in Tab.~\ref{t:spec}, 
Fig.~\ref{f:spec}b). Compared to \cite{fwh03} the fit with the
single Gaussian gives a higher line energy, $E_1 = 390$\,eV vs.~300\,eV, and
smaller line width, $\sigma_1 = 60$\,eV vs.~100\,eV. Part of this change is
due to the improved signal/noise of the spectrum which allows a free fit of
$\sigma_1$ instead of fixing it at 100\,eV as in \cite{fwh03}. 
The updated calibration of the detector might also helpful.
After inclusion of a second Gaussian a successful fit showing no systematic 
residuals was achieved ($\chi^2_\nu = 1.11$ for 1244 degrees of freedom, model
3, Fig.~\ref{f:spec}c).

Since the second Gaussian could be the higher harmonic of a fundamental at
lower energies (\cite{fwh03} proposed an interpretation as proton cyclotron
line in a field of $(2-6) \times 10^{13}$\,G) we made some attempts to fit
the two Gaussians with related parameters. In our model 3, both Gaussians were
fitted independently, they were centered at line energies $E_{\rm 2} \simeq 2
\times E_{\rm 1}$. Hence, if the parameters are forced to obey this
relation, the achieved fit becomes equally good and the other parameters 
are found almost unchanged (model 4). The picture changes slightly if the
width of the two lines is pre-set to the same, adjustable value (model 5). 
This broadens the width of the principal first line 
slightly from 0.15\,keV to 0.17\,keV and shifts the line energy of the second
Gaussian from 0.46\,keV to 0.73\,keV at the same quality of the fit
($\chi^2_\nu =1.11$). Nevertheless the flux in the second Gaussian is reduced
by 50\% in this fit. The last model listed in Tab.~\ref{t:spec} (number 6)
uses a different relation between the line energies, $E_2 = 1.5 \times E_1$,
and was similarly successful with almost unchanged parameters 
compared to the free fit. 

For models 1 and 2 no formal parameter errors were determined because
those models did not adequately describe the spectrum. 
The formal errors of the parameters of fit 3 are: 
$\Delta N_{\rm H} =0.2 \times 10^{20}$\,cm$^{-2}$, $\Delta kT = 2$\,eV, 
$\Delta E_1 = 0.02$\,keV, 
$\Delta \sigma_1 = 0.02$\,keV, 
$\Delta F_1 = 4 \times 10^{-4}$\,keV cm$^{-2}$ \crat,
$\Delta E_2 = 0.03$\,keV, 
$\Delta \sigma_2 = 0.03$\,keV, and
$\Delta F_2 = 4\times 10^{-4}$\,keV cm$^{-2}$ \crat. 
The above experiments have shown cross-talk between the
parameters, hence the true uncertainties are larger than the mentioned
statistical ones. For example, 
models 3 -- 6 are indistinguishable on statistical
grounds but some of their parameters are rather different. 
The main problem consists in a proper location of the continuum. 
The primary Gaussian is so broad, that the low-energy end of the spectrum is
still affected by this feature. This allows the flux in this features 
to be traded
against the amount of interstellar absorption and the temperature of the
blackbody. At the high-energy end of the spectrum photons are detected up to
2\,keV which in principle gives a better leverage for constraining the
blackbody temperature. Our fitted model assumes that these photons belong to
the Wien tail of the blackbody. Consequently, the newly determined blackbody
temperature is higher than found before from a spectrum which missed those
high-energy photons and was affected by the structure which is now described
by a second Gaussian in its presumed Wien-tail. 

\section{Results and discussion}
The joint analysis of six observations of RBS1223 with XMM-Newton has led to
a slight revision of the spectral parameters. The spectrum can be described
with a  
blackbody and two Gaussian absorption lines superimposed. The blackbody
temperature is significantly higher than previously found thanks to the
detection of X-ray photons up to 2\,keV. Also the parameters of the primary
Gaussian absorption line are changed: it appears significantly narrower and
more blueshifted than before. Whether we have seen the true continuum at any
energy remains an open question. 

Similarly unclear is the nature of the features which were parameterised
with two Gaussian absorption lines. They have a flux ratio of about 5:1,
their equivalent widths of EW$_1 \sim 200$\,eV (for $E_1 =0.23$\,keV) 
and EW$_2 \sim 180$\,eV (for $E_2 = 0.46$\,keV) are rather large. 
Their relative line energy 
and the rather large width of the lines are supportive of the
proton cyclotron interpretation in a field of few times $10^{13}$\,G (the line
center of the primary and hence the derived magnetic field strength is not
well contrained, in model 5 the best fit is achieved for $E_1 = 0.2$\,keV,
i.e.~it converges to the lower bound of the pre-set parameter range),
consistent with the observed spin down rate of the star. The rather 
high flux ratio of the two lines, nevertheless, makes an interpretation of the two
lines as being harmonics of each other unlikely, since the oscillator strength
of the first harmonic is smaller than that of the fundamental by a factor
$\sim E/(mc^2)$, which becomes very small for proton masses \cite{pav80}.
An alternative could be to associate the two lines with different fields from
e.g.~the two polar caps. Since the field was shown to be somehow non-dipolar
or off-centered at least \cite{asea05}, different field strength could be
encountered at the two caps. 
It is difficult to test this hypothesis on the basis of the current
data and it needs to be seen if the results of the phase-resolved spectroscopy
reveal any helpful indication in this respect. However, if the two lines would
be the fundamentals of two different cyclotron systems from two different
regions, the infered field strengths would be $B_1 = 4.2 \times 10^{13}$\,G
and $B_2 = 8.5 \times 10^{13}$\,G according to $B (10^{13}{\mbox G}) =
\frac{1}{1.16} (1+z) \frac{m_p}{m_e} E(\mbox{keV})$ and assumed $z=0.18$ (see
\cite{asea05}), $E_1 = 0.23$\,keV and $E_2 = 0.46$\,keV. Hence, 
one field would be slightly below and one above the critical quantum field, where
vacuum polarization becomes important in shaping the lines (see \cite{laiho03}
\cite{laiho04}, and the discussion in \cite{vkea04}). Since the second
Gaussian line (higher field) is rather broader than the primary Gaussian,
this scenario seems to be unlikely.

A possibility which needs to be explored in detail is a blend of
magnetically shifted atomic transitions with proton cyclotron resonances. As
the model computations of \cite{laiho04} indicate, the amount of ionisation in
the assumed hydrogen atmosphere will play a crucial role. An as accurate as
possible temperature determination is thus very important for further
progress. Due to its photometric stability on the one hand and the pronounced
spin variability on the other hand, RBS1223 is a very promising target to gain
further insight from even deeper X-ray spectroscopy.

\end{document}